\documentclass[11pt]{aastex}
\usepackage{graphicx,emulateapj5,apjfonts}
\usepackage{onecolfloat}
\usepackage{epsf}

\shortauthors{Bell et al.}
\shorttitle{Distant red sequence galaxy morphologies}

\newcommand{\combo}{{\sc combo-17 }}

\slugcomment{{\sc ApJ Letters, in press: } 6th November 2003 }

\begin{document}


\def\head{

\title{GEMS Imaging of Red Sequence Galaxies at $z \sim 0.7$: Dusty or Old?}

\author{Eric F.\ Bell$^1$, 
Daniel H.\ McIntosh$^2$, Marco Barden$^1$, Christian Wolf$^3$, 
John A.\ R.\ Caldwell$^4$, Hans-Walter Rix$^1$, 
Steven V.\ W.\ Beckwith$^4$,
Andrea Borch$^1$, Boris H\"aussler$^1$, Knud Jahnke$^5$, 
Shardha Jogee$^4$, 
Klaus Meisenheimer$^1$, Chien Y.\ Peng$^6$, Sebastian F.\ Sanchez$^5$,
Rachel Somerville$^4$, Lutz Wisotzki$^5$}
\affil{$^1$ Max-Planck-Institut f\"ur Astronomie,
K\"onigstuhl 17, D-69117 Heidelberg, Germany; \texttt{bell@mpia.de}\\
$^2$ Department of Astronomy, University of Massachusetts, 
710 North Pleasant Street, Amherst, MA 
01003, USA \\ 
$^3$ Department of Physics, Denys Wilkinson Bldg., University
of Oxford, Keble Road, Oxford, OX1 3RH, UK \\
$^4$ Space Telescope Science Institute, 3700 San Martin Drive, Baltimore MD, 21218, USA \\
$^5$ Astrophysikalisches Institut Potsdam, An der Sternwarte 16, D-14482 Potsdam, Germany \\
$^6$ Steward Observatory, University of Arizona, 933 N.\ Cherry
Ave., Tucson AZ, 85721, USA 
}

\begin{abstract}
We have used the $30'\times 30'$ {\it Hubble Space Telescope}
image mosaic from the Galaxy Evolution from Morphology and 
SEDs ({\sc gems}) project in conjunction with the \combo 
deep photometric redshift survey to define 
a sample of nearly 1500 galaxies with $0.65 \le z \le 0.75$.
With this sample, we can study the distribution of rest-frame 
$V$-band morphologies more than 6 Gyr ago, without differential 
bandpass shifting and surface brightness dimming across this 
narrow redshift slice.
Focusing on red-sequence galaxies at $z \sim 0.7$, we
find that 85\% of their combined rest-frame $V$-band luminosity density
comes from visually-classified E/S0/Sa galaxies down
to $M_V - 5\log_{10}h \la -19.5$.  Similar results
are obtained if automated classifiers are used.  This fraction 
is identical to that found at the present day, and is biased by less
than 10\% by large scale structure and the morphology--density
relation.  Under the assumption that peculiar and edge-on disk
galaxies are red by virtue of their dust content, 
we find that less than 13\% of the total rest-frame $V$-band
luminosity of the $z \sim 0.7$ red galaxy population is from dusty galaxies.
\end{abstract}

\keywords{galaxies: general ---  
galaxies: evolution --- galaxies: stellar content ---
galaxies: elliptical and lenticular }
}

\twocolumn[\head]

\section{Introduction}

A key discriminant between hierarchical 
galaxy formation in a Cold Dark Matter
(CDM) Universe and monolithic galaxy formation 
scenarios is the evolution of the number
of spheroid-dominated galaxies 
\citep[e.g.,][]{kauffmann96,aragon98}.  Yet,  
it is impossible to identify distant spheroid-dominated galaxies reliably
and measure this evolution without sub-arc\-sec\-ond resolution imaging
of large numbers of galaxies with known redshifts.  The largest
representative sample of 
morphologically early-type galaxies
studied to date was a {\it HST} study of 150 galaxies 
\citep{im02}, which was too small to
differentiate strongly between hierarchical and monolithic 
formation of the early-type population.  To collect larger
samples, workers have instead photometrically selected the reddest
galaxies, noting that morphologically early-type galaxies in the local 
Universe and distant clusters populate a well-defined red sequence  
\citep[e.g.,][]{ble92,schweizer92,hogg02,kodama97}.
In this {\it Letter}, we explore the {\it HST} rest-frame optical $V$-band
morphologies of 1492 galaxies in the thin
redshift slice $0.65 \le z \le 0.75$
to understand how well the evolution of 
red galaxies reflects that of the early-type 
galaxy population. 

Careful study of color-selected galaxies has yielded
impressive results\footnote{Other groups have chosen to study
the evolution of distant massive $K$-band selected galaxies.
This is a roughly equivalent test, as the most massive galaxies are 
red at redshifts out to at least one \citep{bell03}.  
In common with the more recent
color-selected surveys, they find mild departures from 
passive evolution, conceptually consistent with galaxy
formation in a hierarchical Universe 
\citep[e.g.,][]{kauffmann98,drory01,somerville03}.}.  
Previous work \citep[e.g.,][]{lilly95,lin99,pozetti03}
found little evolution in the number density of red galaxies,
but with error bars large enough to be consistent
with both monolithic and hierarchical evolution 
of the red galaxy population.  Recently, \citet{chen03}
and \citet{bell03} have used samples of 5000 and 25000 galaxies
with high-quality photometric redshifts to 
show that the total stellar mass in the red galaxy population
has increased monotonically since
$z \sim 1$.  In the latter case, the sample 
size was large enough to
rule out purely passive evolution with 
high significance,
favoring instead a factor of two stellar mass evolution, 
in excellent qualitative and quantitative agreement
with hierarchical models of galaxy formation.   

Nevertheless, a significant or even dominant fraction 
of distant $z \ga 0.5$ red galaxies may be red owing to their dust
contents, rather than because their stellar populations are old.
In the local Universe, typically $\sim 3/4$
of the galaxies on the red sequence are morphologically early-type
\citep[E, S0 or Sa;][]{strateva01,hogg02}.  We illustrate this
in panel {\it a)} of Fig.\ \ref{fig:cmr}, 
where we show evolution and $k$-corrected
$U-V$ colors as a function of absolute magnitude for a 
$1/V_{\rm max}$ selected\footnote{
Where $V_{\rm max}$ is the volume within which one could observe 
each galaxy, making this sample essentially volume-limited down to 
a limiting magnitude.} sample of 1500 galaxies from the 
Sloan Digital Sky Survey (SDSS) selected to have $M_V -5\log_{10}h < -18$
\citep{sdss,bell03}.  
Open blue symbols show
morphologically late-type galaxies, and solid red symbols 
show morphologically
early-type galaxies.  We find that 82\% of the combined $V$-band
luminosity density from red sequence galaxies in the local Universe
comes from morphologically early-type galaxies.  Yet, 
extremely red objects (EROs), galaxies with colors
characteristic of early-type galaxies at 
$1 \la z \la 2$, seem to be a mix of dusty
star-forming galaxies, edge-on spiral galaxies and
early-type galaxies, with perhaps as little
as 30\% of the ERO population being red by virtue of 
their old stellar populations \citep[e.g.,][]{yan03,moustakas03}.
{\it Therefore, color-selected samples may not give a true 
picture of the evolution of early-type galaxies}.

To better connect between these new large studies of the redshift evolution
of the red galaxy population and the evolution of morphologically
early-type galaxies, it is necessary to explore 
the rest-frame optical morphology of red galaxies at redshifts closer to unity.
In this {\it Letter}, we explore the rest-frame $V$-band
morphologies of nearly 1500 galaxies with both deep F850LP
{\it HST} data from the Galaxy Evolution from Morphology and
SEDs ({\sc gems}) survey (Rix et al., in prep.) and accurate
photometric redshifts in the thin redshift slice 
$0.65 \le z \le 0.75$ from the \combo survey
\citep[`Classifying Objects by Medium-Band 
Observations in 17 Filters';][]{wolf03}.
Throughout, we assume
$\Omega_{\rm m} = 0.3$, $\Omega_{\rm m}+\Omega_{\Lambda} = 1$, and 
$H_0 = 100 h $\,km\,s$^{-1}$\,Mpc$^{-1}$ (where $h = 0.7$) following 
\citet{spergel03}.
 
\section{The Data} \label{sec:data}

To date, \combo has surveyed three disjoint
$\sim 34' \times 33'$ southern and equatorial fields
to deep limits (complete to $m_R \sim 24$, with integration times of 
$\sim 180$\,ksec/field)
in 5 broad and 12 medium passbands.
Using these deep data in conjunction with 
non-evolving galaxy, star, and AGN template spectra, objects
are classified and redshifts assigned for $\sim 99$\% of the
objects.  Typical galaxy redshift accuracy
is $\delta z/(1+z) \sim 0.02$ \citep[Wolf et al., in preparation;][]{wolf03},
allowing construction of $\sim 0.1$ mag accurate 
rest-frame colors and absolute magnitudes.

To explore galaxy morphology in the rest-frame
optical from a single observed passband, we
study galaxies in one thin redshift slice.   
Here, we select galaxies from \combo in the 
Extended Chandra Deep Field South (E-CDFS)
for morphological classification in the narrow interval
$0.65 \le z \le 0.75$ (corresponding to $\Delta t \sim 0.5$\,Gyr, minimizing
galaxy evolution across this slice).  
At this redshift, F850LP samples rest-frame
$\sim V$-band, allowing comparison with 
local samples.
Furthermore, the sample size is maximized by 
being at the peak of the \combo number counts and through our inclusion of 
the only two major large-scale structures in the E-CDFS, both 
at $z \sim 0.7$ \citep{gilli03}.

We use F850LP imaging from the {\sc gems}
survey (Rix et al., in prep.) to provide 
sub-arcsecond resolution rest-frame $V$-band data for 
our sample of $0.65 \le z \le 0.75$ galaxies.  
{\sc gems} surveys a $\sim 30' \times 30'$
portion of the E-CDFS in the F606W and F850LP passbands
to deep limits using the Advanced Camera for Surveys \citep{ford03}
on the {\it HST}.  The {\sc gems} area is covered by a multiple, overlapping
image mosaic that includes the smaller but deeper {\sc goods}
area \citep{giavalisco03}.  One orbit per pointing
was spent on each passband (63 {\sc gems} tiles and 15 {\sc goods}
tiles), allowing galaxy detection to a limiting surface 
brightness of $\mu_{\rm F850LP,AB} \sim 24$\,mag\,arcsec$^{-2}$
(H\"au\ss ler et al., in prep).  At $z = 0.7$, the ACS resolution 
of $\sim 0.05''$ corresponds to $\sim 350$\,pc resolution, 
roughly equivalent to $\sim 1''$ resolution at Coma
cluster distances.

\begin{figure*}[th]
\epsfxsize 18cm
\epsfbox{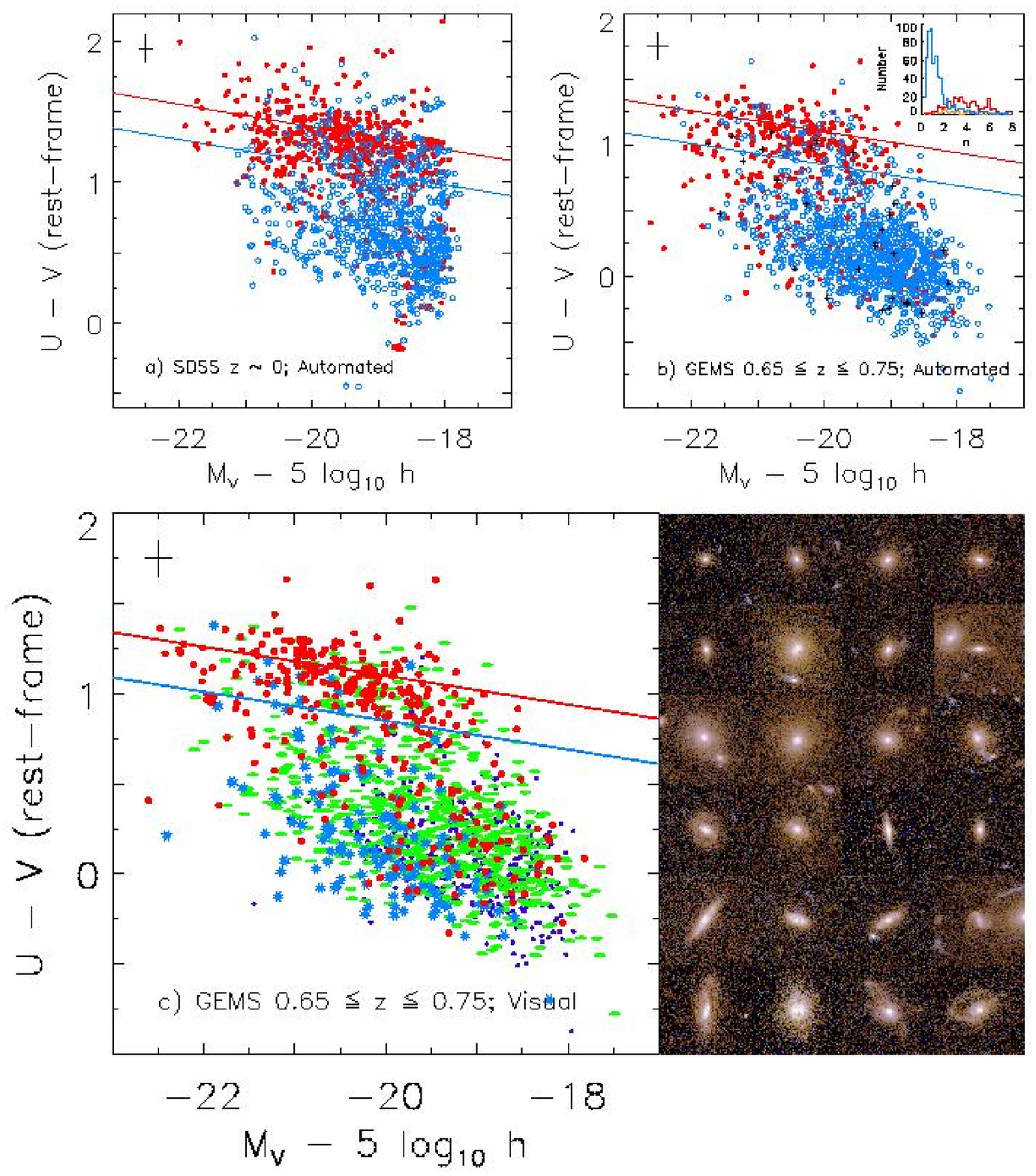}
\caption{\label{fig:cmr} The morphological types
of galaxies as a function of 
their rest-frame $V$-band absolute magnitude and 
$U-V$ color over the last half of cosmic history.
{\it a)} The colors and magnitudes of 1500 
morphologically-classified local Universe 
galaxies from the SDSS.  
Blue open circles denote morphologically-classified late-type galaxies
($c_r < 2.6$) whereas red solid symbols show early types 
\protect\cite[$c_r \ge 2.6$; see, e.g.,][for more discussion of 
$c_r$ as a morphological indicator]{strateva01,colblanton}.
The red line shows a fit to the red sequence, and the blue line
the adopted cut between red sequence and blue galaxies.  $U-V$ colors
and $M_V$ values were synthesized using the observed $ugr$ magnitudes
for a $14 < r < 16.5$ apparent magnitude limited sample, from which 
galaxies were randomly drawn weighted by $1/V_{\rm max}$ to approximate
a volume limited sample of galaxies.  Only galaxies 
with $M_V - 5 \log_{10}h < -18$ were chosen to 
approximate the magnitude limit of the $z=0.7$ sample
of 1492{ }\combo galaxies. {\it b)} The morphologies of 
GEMS $0.65 \le z \le 0.75$ galaxies, using the S\'ersic 
$n$ parameter as the morphological classifier.  Filled red circles
denote early-type galaxies with $n \ge 2.5$, and open blue circles
late-types with $n < 2.5$.  Black crosses denote galaxies which
were not successfully fit by either {\sc galfit} or {\sc gim2d}.
The solid lines denote fits
to the red sequence (red) and the adopted separation between 
red and blue galaxies (blue), as in panel {\it a)}.  The inset
panel shows the $n$ distributions of the visually-classified
E/S0 galaxies (red), Sb-Sdm galaxies (blue), and Sa galaxies
(yellow).  A cut at $n=2.5$ clearly 
separates early-type and late-type galaxies with 80\% reliability and
less than 25\% contamination.  The inclusion of Sa galaxies 
in either of
the late or early-type galaxy bins does not significantly affect
this conclusion.
{\it c)} The visual classifications of the GEMS
$z \sim 0.7$ galaxies.  Red circles denote visually-classified 
E/S0 galaxies, green ellipses are Sa--Sm galaxies, 
blue stars are Peculiar/Strong Interaction galaxies, 
purple circles are Irregular/Weak Interaction and compact
galaxies.  We show also color postage stamps 
of red sequence galaxies: the top three lines are
visually classified as E or S0 (where we show an unrepresentatively 
large fraction of E/S0s with substructure), the next two lines are 
classified as Sa--Sm, and the final line of galaxies are classified
as Peculiar/Strong Interaction.  Postage stamps are 6$''$ square,
corresponding to 35\,kpc in the adopted cosmology.
} \vspace{-0.4cm}
\end{figure*}

Galaxy classification was carried out both by-eye
and automatically on our final 
sample of 1492 galaxies at $0.65 \le z \le 0.75$.  We adopt 
by-eye classification bins of 
E/S0, Sa, Sb--Sdm, Peculiar/Strong Interaction, 
Irregular/Weak Interaction, and 
unclassifiable (typically because of small galaxy size).
Galaxies were classified on the basis of both central light concentration
and smoothness.  In particular, the difference between S0 and Sa is
largely a difference in smoothness, whereas the distinction between
Sa and Sb is largely driven by central concentration\footnote{Note, however,
that our qualitative conclusions remain unchanged whether or not
we class the Sa galaxies with the spirals or E/S0s.}.
The Peculiar/Strong Interaction designation includes galaxies with
what appeared to be tidal tails and/or multiple nuclei; below a
certain magnitude cut, these features will be too faint to 
readily recognize, and the galaxies will be classified as 
Irregular/Weak Interaction.  
For automated galaxy classification, 
we fit single-component S\'{e}rsic models to all 1492 galaxies using
the {\sc galfit} \citep{peng02} and {\sc gim2d} \citep{simard02} 
software packages.  The
\citet{sersic68} model has a profile with
surface brightness $\Sigma \propto e^{-r^{1/n}}$, where $r$ is the 
radius and $n$ is an index denoting how 
concentrated the light distribution of a given galaxy is:
$n = 1$ corresponds to an exponential light distribution, and 
$n = 4$ corresponds to the well-known de Vaucouleurs profile.
Typical uncertainties are $\delta n/n \sim 0.2$ for $\ga 90\%$
of galaxies, as estimated using
simulations and {\sc galfit/gim2d} inter-comparisons.  
In what follows, we adopt
$n$ values from {\sc galfit}, using {\sc gim2d} to fill in values
for galaxies without successful {\sc galfit} fits.  Less than 
2\% of galaxies were not successfully fit by either code; in 
panel {\it b)} of Fig.\ \ref{fig:cmr} these galaxies are denoted by
black crosses.

We compare the automated and visual classifications in the inset panel 
of panel {\it b)} of Fig.\ \ref{fig:cmr},
where we show the $n$ distribution of Sb--Sm galaxies as the
blue histogram, and the $n$ distribution of E/S0 galaxies 
as the red histogram.  One can immediately see that visually-classified
spiral galaxies prefer $n \sim 1$, whereas visually-classified early-type
galaxies have higher $n$ values, broadly distributed around $n \sim 4$.  
In particular, we find that roughly 80\% of 
the visually-classified early-type galaxies have $n \ge 2.5$, with
less than 25\% contamination from later types.
This success rate is very similar to the success rate achieved
by the Sloan collaboration
using a $r$-band concentration parameter cut.

\section{The Morphologies of $z\sim 0.7$ red sequence galaxies} 
 \label{sec:results}

In panels {\it b)} and {\it c)} of Fig.\ \ref{fig:cmr}, 
we show the automatically and visually-determined
morphological types of $0.65 \ge z \ge 0.75$ galaxies as a function of their 
rest-frame $V$-band absolute magnitude and rest-frame
$U-V$ color.  These particular passbands are chosen for
consistency with classic studies of the colors of 
galaxies in the local Universe \citep[e.g.,][]{sandage78,ble92}, 
and with the color-based study of 
\citet{bell03}.  

Focusing on the red sequence galaxies (those redder than the blue
line) at $z \sim 0.7$, it is immediately clear that the majority of these
galaxies are morphologically early-type (see also the color
postage stamps of example red-sequence galaxies)\footnote{We
defer a study of the nature of blue, morphologically-compact
galaxies to a later date.}.  Down to 
$M_V -5\log_{10}h \sim -19.5$,
we find that 74\% of the rest-frame $V$-band luminosity
density on the red sequence is from E/S0 galaxies, rising to 85\%
if Sa galaxies are included as morphologically early-type.  Automatic
classification gives similar results, with 78\% of the $V$-band
luminosity density coming from galaxies with Sersic indices $n \ge 2.5$.
These fractions
are very similar to the local Universe (82\% of the $V$-band luminosity
density; see \S 1) and with an E/S0 fraction of $\sim 60$\% for the
$z>0.6$ red galaxy subsample from \citet{im02}.
The remaining $z \sim 0.7$ red sequence luminosity density comes from spiral
galaxies (10\%), which are primarily highly inclined, 
and interacting/peculiar galaxies (5\%).  

Furthermore, under the reasonable assumption that galaxies that are red owing
to their dust content will have peculiar (5\% of the total rest-frame
$V$-band luminosity
density) and/or highly-inclined disk (8\% of the total luminosity
density) morphologies, we place an upper limit of 13\% on the fraction of 
rest-frame $V$-band luminosity density that is from galaxies
that are dust-reddened rather than old.  Interestingly,
most dust-reddened galaxies at $z \sim 0.7$ are edge-on spiral
galaxies rather than dust-enshrouded starbursts.

The comoving volume probed by this sample is only
$5 \times 10^4 h^{-3}$ comoving Mpc$^{-3}$.
Furthermore, this redshift range contains
the only significant large-scale structure in the CDFS; therefore,
owing to the morphology-density relation \citep[e.g.,][]{dressler97},
it is possible that the relative fraction of early-type galaxies to the
red galaxy population would be somewhat lower in a 
cosmologically-representative volume.
To check this, we explored the F850LP visual 
morphologies of an unbiased sample of 
51 $0.8 \le z \le 0.9$ galaxies with $M_V - 5\log_{10}h \le -20$
and $U-V \ge 1.0$.  We found that 66\% of the red-sequence
galaxies were E/S0, 12\% were edge-on disks, 14\% were less 
inclined disks (Sa--Sb), and 8\% were possible interactions.  
Not withstanding the small number statistics, this suggests
that the environmental bias in the fraction of $0.65 \le z \le 0.75$
red galaxies that are morphologically early-type is $\la 10$\%
(74\% E/S0 at $z \sim 0.7$ vs.\ 66\% E/S0 at
$z \sim 0.85$).

\section{Comparison with EROs} \label{sec:disc}

Morphologically early-type galaxies form less than 40\%
of the ERO (galaxies with the colors of 
passively-evolving stellar populations with $1 \la z \la 2$)
population; the rest consists of edge-on disks and 
dusty starbursting galaxies \citep{yan03,moustakas03}.   
Nevertheless, we find that in
complete samples with well-defined redshift ranges $\ga 80$\%
of red sequence galaxies with $M_V -5\log_{10}h \la -19.5$
are morphologically early-type over the last
half of the Universe's evolution.  While we 
cannot rule out strong evolution of the red galaxy population in 
the 3 Gyr between $z \sim 1.5$ and $z \sim 0.7$, it is possible that
the main difference between our and the ERO results is sample selection.
EROs are selected to have observed frame red 
optical--near-infrared colors, and 
are therefore an inhomogeneous mix of intrinsically red 
galaxies at a variety of different redshifts.  For example, 
intrinsically very red but relatively faint and numerous edge-on galaxies
at lower redshifts may have the same apparent $K$-band magnitudes
as rather less intrinsically red but more luminous and rare early-type
galaxies at higher redshifts, boosting the observed fraction of edge-on
galaxies.   
A detailed census of a large sample of EROs with 
morphological {\it and} redshift information should help to definitively
disentangle the roles of dust and old stellar populations in driving
the colors of red galaxies at $z \ga 1$.

\section{Monolithic or hierarchical formation of spheroids?}

We find that the dominant portion of the rest-frame $V$-band
light from red-sequence galaxies at all redshifts $0 < z < 0.75$ 
comes from morphologically early-type galaxies.
At first sight, this result cannot distinguish between a monolithic,
early origin for spheroids or a hierarchical, extended build up.
Yet, taken together with the clear detection of stellar mass 
increase in the red sequence over the last $\sim 10$ 
Gyr \citep{bell03,chen03}, a hierarchical origin 
of spheroids is favored at this stage.  A more detailed
investigation of morphologically-selected
spheroids, decomposing galaxies into both bulge and disk
components, is clearly required to make further
progress towards this important goal. 

\acknowledgements
We wish to thank Myungshin Im and the referee, Alan Dressler for
their constructive comments.
Based on observations taken with the NASA/ESA {\it Hubble Space
Telescope}, which is operated by the Association of Universities
for Research in Astronomy, Inc.\ (AURA) under NASA contract NAS5-26555.
E.\ F.\ B.\ acknowledges the financial support provided through the European
Community's Human Potential Program under contract 
HPRN-CT-2002-00316, SISCO.
D.\ H.\ M.\ acknowledges support
by JPL/NASA through the 2MASS core science projects.
C.\ W.\ was supported by the PPARC rolling grant in Observational
Cosmology at the University of Oxford.
S.\ F.\ S.\ is supported by the Euro3D European RTN, under
contract HPRN-CT-2002-00305.
This publication makes use of the 
{\it Sloan Digital Sky Survey} (\texttt{http://www.sdss.org/}).

\end{document}